\documentclass[journal,10pt]{IEEEtran}
\usepackage{graphicx}
\usepackage{amsfonts}
\usepackage{amsmath}
\usepackage{amssymb}
\usepackage{bm}
\usepackage{color}
\usepackage{graphicx,subfigure}
\usepackage{amssymb}
\usepackage{times}
\usepackage{latexsym}
\usepackage{bm}
\usepackage{cases}
\usepackage{array}
\usepackage{subfigure}
\usepackage{url}
\usepackage{algorithm}
\usepackage{algorithmic}
\usepackage{multirow}
\usepackage{epstopdf}
\usepackage{epsfig}
\usepackage{fancybox}
\usepackage{textcomp}

\usepackage{cite}
\usepackage{autobreak}

\begin{document}



\title{Hybrid Beamforming for RIS-Empowered Multi-hop Terahertz Communications: A DRL-based Method}

\author{Chongwen~Huang,~\IEEEmembership{Member,~IEEE,}  Zhaohui~Yang,~\IEEEmembership{Member,~IEEE,} George~C.~Alexandropoulos,~\IEEEmembership{Senior Member,~IEEE}, Kai Xiong, Li Wei Chau~Yuen,~\IEEEmembership{Senior Member,~IEEE}, and Zhaoyang Zhang,~\IEEEmembership{Member,~IEEE} 

%
}

\maketitle

\begin{abstract}

Wireless communication in the TeraHertz band (0.1--10 THz) is envisioned as one of the key enabling technologies for the future six generation (6G) wireless communication systems. However, very high propagation attenuations and molecular absorptions of THz frequencies often limit the signal transmission distance and coverage range. Benefited from the recent breakthrough on the reconfigurable intelligent surfaces (RIS) for realizing smart radio propagation environment, we propose a novel hybrid beamforming scheme for the multi-hop RIS-assisted communication networks to improve the coverage range at THz-band frequencies. We investigate the joint design of digital beamforming matrix at the BS and analog beamforming matrices at the RISs, by leveraging the recent advances in deep reinforcement learning (DRL) to combat the propagation loss. Simulation results show that our proposed scheme is able to improve 50\% more coverage range of THz communications compared with the benchmarks. Furthermore, it is also shown that our proposed DRL-based method is a state-of-the-art method to solve the NP-bard beamforming problem, especially when the signals at RIS-empowered THz communication networks experience multiple hops.
\end{abstract}


\vspace{-0.25cm}
\section{Introduction}

Future sixth generation (6G) wireless communication systems are expected to rapidly evolve towards an  ultra-high speed and low latency with the software-based functionality paradigm \cite{Akyildiz_Thz2018wcm,han2014multi,Sarieddeen2020wcm,chongwenhmimos,MarcoJSAC2020}.  Although current millimeter-wave (mmWave) communication systems (30-300 GHz) have been intergraded into 5G mobile systems, and several mmWave sub-bands were released for licensed communications, e.g., 27.5-29.5 GHz, 57-64 GHz, 81-86 GHz, etc., the total consecutive available bandwidth is still less than 10GHz, which is difficult to offer Tbps data rates \cite{Akyildiz_Thz2018wcm,han2014multi,Sarieddeen2020wcm,Moldovan2016}. To meet the increasing demand for higher data rates and new spectral bands, the Terahertz (0.1--10 THz) band communication is considered as one of the promising technology to enable ultra-high speed and low-latency communications. Although  major progresses in the recent ten years are empowering practical THz communication networks,  there are still  many challenges in THz communications that require innovative solutions. One of the major challenges is the very high propagation attenuations, which drastically reduces the propagation distance.

Fortunately, the recently proposed reconfigurable intelligent surface (RIS) is considered as a promising technology to combat the propagation distance problem, since RIS can be programmed to change an impinging electromagnetic (EM) field in a desired way to focus, steer, and enhance the signal power towards the object user \cite{hcw2018icassp,MarcoJSAC2020,chongwenhmimos,zappone2020overhead,chongwentwc2019,wqq2019beamforming,
Basar2019access}. Recently, RIS-based designs have emerged as strong candidates that empower communications at the THz band. Specifically, \cite{Akyildiz_Thz2018wcm,Sarieddeen2020wcm} presented some promising visions and potential applications leveraging the advances of RIS to combat the propagation attenuations and molecular absorptions of THz frequencies. To remove obstacles of realizing these applications, \cite{ning2019beamforming,weili_CE_2020} proposed the channel estimation and data rate
maximization transmission solutions for massive multiple input multiple output (MIMO) RIS-assisted THz system. Furthermore, some beamforming and resource allocation schemes were proposed in \cite{ning2019beamforming,Nie2019resource}. For example,  a cooperative beam training scheme and two cost-efficient hybrid beamforming schemes were proposed in \cite{ning2019beamforming} for the THz multi-user massive MIMO system with RIS, while a resource allocation based on the proposed end-to-end physical model was introduced in \cite{Nie2019resource} to  improve the achievable distance and data-rate at THz band RIS-assisted communications.

All above works assume single-hop RIS assisted systems, where only one RIS is deployed between the BS and the users. In practical, similar to multi-hop relaying systems, multiple RISs can be used to overcome severe signal blockage between the BS and users to achieve better service coverage. Although multi-hop MIMO relaying systems have been addressed in the literature intensively in the context of relay selection, relay deployment, and precoding design, multi-hop RIS assisted systems have not yet been studied. In addition, the methodologies developed for multi-hop relay systems cannot be directly applied to multi-hop RIS assisted systems, due to different reflecting mechanisms and channel models. Particularly, the constraint on diagonal phase shift matrix and unit modulus of the reflecting RIS makes the joint design of transmit beamforming and phase shifts extremely challenging.

To address high-dimension, complex EM environment, and mathematically intractable non-linear issues of communication systems, the model-free machine learning method as an extraordinarily remarkable technology has introduced in recent years \cite{chen2019FLwireless,Ref14c}.  Overwhelming research interests and results uncovers  machine learning technology to be used in the future 6G wireless communication systems for dealing with the non-trivial problems due to extremely large dimension in large scale MIMO systems. To be specific, deep learning has been used to obtain the channel state information (CSI) or beamforming matrix in non-linear communication systems. In terms of dynamic and mobile wireless scenarios, deep reinforcement learning (DRL) provides an effective solution by leveraging the advantages of deep learning, iterative updating and interacting with environments over the time \cite{Ref14c,Ref16e,Ref16g,deeplearn2},. In particular, the hybrid beamforming matrices were obtained by DRL for the mobile mmWave systems in \cite{Ref16e}, while  \cite{Ref16g} proposed a novel idea to utilize DRL for optimizing the network coverage.

In this paper, we present a multi-hop RIS-assisted communication scheme to overcome the severe propagation attenuations and improve the coverage range at THz-band frequencies, where the hybrid design of transmit beamforming at the BS and phase shift matrices is obtained by the advances of DRL. Specifically, benefited from the recent breakthrough on RIS, our main objective is to overcome propagation attenuations at THz-band communications by deploying multiple passive RISs  between the BS and multiple users. To maximize the sum rate, formulated optimization problem is non-convex due to the multiuser interference, mathematically intractable multi-hop signals, and non-linear constraints. Owning to the presence of possible multi-hop propagation, which results in composite channel fadings, the optimal solution is unknown in general. To tackle this intractable issue, a DRL based algorithm is proposed to find the feasible solutions.

The notations of this paper are summarized as follows. We use the $\mathbf{H}$ to denote a general matrix.  $\mathbf{H}^{(t)}$ is the value of $\mathbf{H}$ at time $t$. $\mathbf{H}^T$, and $\mathbf{H}^{\mathcal{H}}$ denote the transpose and conjugate transpose of matrix $\mathbf{H}$, respectively. $Tr \{ \}$ is the trace of the enclosed. For any vector $\mathbf{g}$, $\mathbf{g}(i)$ is the $i^{th}$ entry, while $\mathbf{g}_{k}$ is the channel vector for the $k^{th}$ user. $||\mathbf{h}||$ denotes the magnitude of the vector. $\mathcal{E}$ denotes statistical expectation. $|x|$ denotes the absolute value of a complex number ${x}$, and its real part and imaginary part are denoted by $Re{(x)}$ and $Im{(x)}$, respectively.

\vspace{-0.15cm}
\section{System Model and Problem Formulation}

\subsection{Terahertz-Band Channel Model}

Unlike the lower frequency band communications, a signal operating at the THz band can be affected easily by many peculiarly factors, mainly is influenced by the molecular absorption due to water vapor and oxygen, which result in very high path loss for line-of-sight (LOS) links\cite{Akyildiz_Thz2018wcm,han2014multi,Sarieddeen2020wcm}. On the other hand, spreading loss also contributes a large proportion of attenuations. In terms of  non-line-of-sight (NLOS) links, besides mentioned peculiarities, unfavorable material and roughness of the reflecting surface also will cause a very severe reflection loss\cite{Moldovan2016,han2014multi,Nie2019resource}. The overall channel transfer function can be written as,

\begin{equation} \label{eq:modelthz1}
\begin{split}
H(f,d,\bm{\zeta})=&H^{LOS}(f,d)e^{-j2\pi f\tau_{LOS}}+ \\
&\sum^{M_{rays}}_{i=1}H_i^{NLOS}(f,\zeta_i)e^{-j2\pi f\tau_{NLOS_i}},
\end{split}
\end{equation}
where $f$ denotes the operating frequency, $d$ is the distance between the transmitter and receiver, the vector $\bm{\zeta}=[\zeta_1,...,\zeta_{M_{rays}}]$ represents the coordinates of all scattering points, and $\tau_{LOS}$ and $\tau_{NLOS_i}$ denote  the propagation delays of the LOS path and $i^{th}$ NLOS path respectively.


\subsection{Proposed Multi-hop Scheme}

\begin{figure*}[t]
\begin{center}
  \includegraphics[width=14cm]{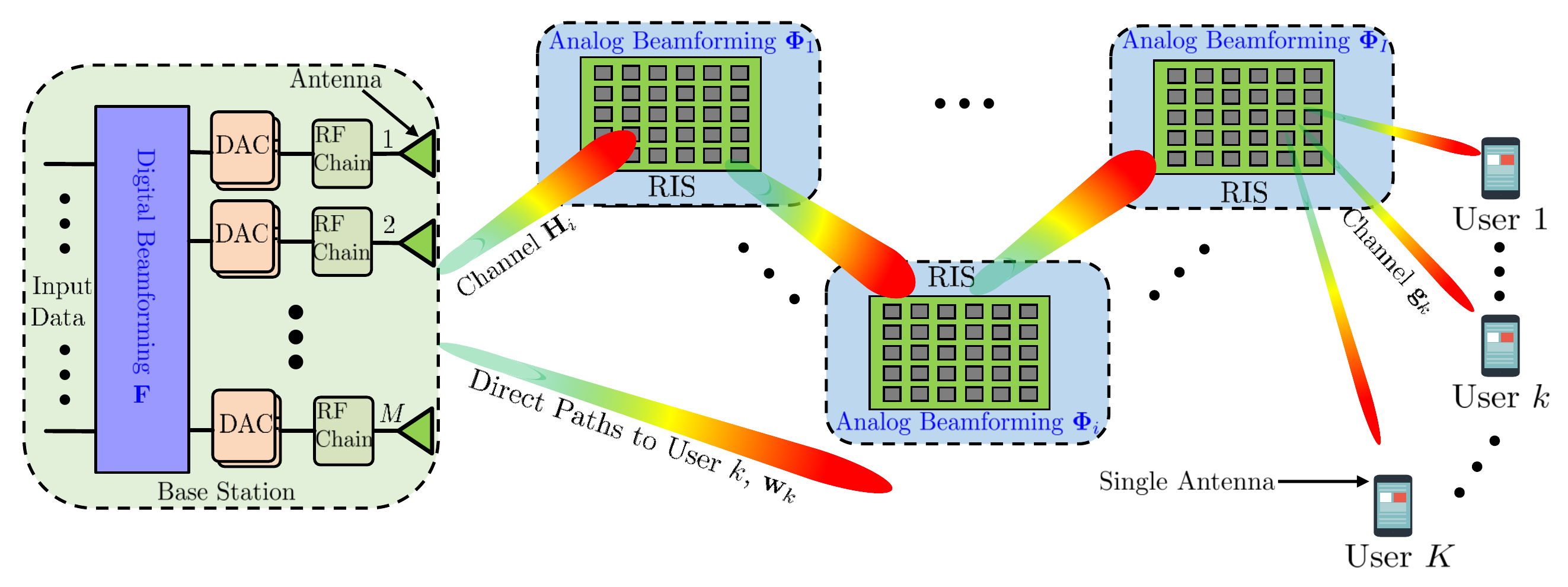}
  \caption{The RIS-based multi-hop for THz communications and proposed practical RIS-based hybrid beamforming architecture.  }
  \label{fig:hybrid}
\end{center}  \vspace{-6mm}
\end{figure*}

As mentioned before, communications over the THz band are very different with the low frequency band communications, the transmitted signal suffers from the severe path attenuations. To address this issue, we introduce a multi-hop multiuser system by leveraging some unique features of RISs, which is comprised of a BS, $N$ reflecting RISs and multiple single-antenna users shown in Fig. \ref{fig:hybrid}. We consider that BS equipped with $M$ antennas communicate with $K$ single-antenna users in a circular region. Assume that the $i^{th}$ reflecting RIS, $i=1,\cdots, N$, has $N_i$ reflecting elements. A number of $K$($K\leq M$) data streams are transmitted simultaneously from the $M$ antennas of the BS with the aid of multiple RISs to improve the coverage range of THz communications. Each data stream is beamforming to one of the $K$ users by the assistance of RISs.

\textbf{Remark}: In contrast to the traditional precoding architectures,  a key novelty of this proposed multi-hop scheme is to take full advantages of RISs with the unique programmable feature as an \textbf{external and portable analog precoder}, i.e., the RIS functions as a reflecting array, equivalent to introduce the analog beamfroming to impinge signals, which not only can remove internal analog precoder at BS that simplifies the architecture and reduces its cost significantly, but also improve the beamforming performance of THz-band communication systems.

We assume that the channel fading is frequency flat, and the transmitted signal experiences $I_k (I_k \leq N)$ hops on RISs to arrive $k^{th}$ user. We denote the channel matrix from the BS to the first reflecting RIS as $\mathbf{H}_1 \in \mathbb{C}^{(N_1 \times M )}$, the channel matrix from the $i^{th}$ RIS to the $(i+1)^{th}$ RIS as $\mathbf{H}_{(i+1)} \in \mathbb{C}^{(N_{(i+1)} \times N_i)}$.  The received signal at the $k^{th}$ user is given as
\begin{equation} \label{eq:sys_1}
\begin{split}
y_k=(\mathbf{g}_{k}^T \prod_{i=1,\cdots, I_k} \mathbf{\Phi}_i \mathbf{H}_{i}+\mathbf{w}_k)\mathbf{x}+n_k \\
\end{split}
\end{equation}
where the vector $\mathbf{g}_{k} \in \mathbb{C}^{(N_{I_k} \times 1)}$ and $\mathbf{w}_k \in \mathbb{C}^{( 1 \times M )} $ denote the channel from the last RIS to the $k^{th}$ user and the direct channel from the BS to user $k$ respectively, $\mathbf{\Phi}_i \triangleq\mathrm{diag}[\theta_{i1},\theta_{i1},\ldots,\theta_{iN_i}] \in \mathbb{C}^{(N_{i} \times N_{i})}$ is the phase shift matrix of the $i^{th}$ RIS, i.e., the $i^{th}$ analog precoding matrix, $\mathbf{x} \in \mathbb{C}^{M \times 1} $ is the transmit vector from the BS, and $n_k$ is the additive white Gaussian noise (AWGN) with the zero mean and $\sigma_n^2$ variance. We further assume that the channel $\mathbf{g}_{k}$, $\mathbf{w}_{k}$, and $\mathbf{H}_{i}$ for all $K$ users are perfectly known at both the BS and all users. Although we admit that obtaining these CSIs are  challenging tasks for RIS-based communication systems, there are already significant methods that are proposed in existing works \cite{ning2019beamforming,weili_CE_2020}. Furthermore, research on the channel estimation is also beyond the scope of this paper. Therefore, we have this assumption. The transmit vector $\mathbf{x}$ can be written as $ \mathbf{x}\triangleq\sum_{k=1}^{K}\mathbf{f}_{k}s_{k} $,
where $\mathbf{f}_{k}\in\mathbb{C}^{M\times 1}$ and $s_{k} \in \mathcal{CN}(0,1)$, i.e., under the assumption of Gaussian signals, denote the beamforming vector and independent user symbols respectively.  The power of the transmit signal from the BS has the constraint,
 $ \mathcal{E}[|\mathbf{x}|^2]=\mathrm{tr}(\mathbf{F}^H\mathbf{F})\leq P_t\;$
wherein $\mathbf{F}\triangleq[\mathbf{f}_1,\mathbf{f}_2,...,\mathbf{f}_K]\in\mathbb{C}^{M\times K}$, and $P_t$ is the total transmission power of the BS.

It should be noted that $\mathbf{\Phi}_i$ is a diagonal matrix whose entries are given by $\mathbf{\Phi}_i(n_i,n_i)=\theta_{in_i}=e^{j\phi_{n_i}}$, where $\phi_{n_i}$ is the phase shift induced by each element of the RIS. Like a mirror, the signal goes through the RIS is no energy loss, which means  $|\mathbf{\Phi}_i(n_i,n_i)|^2$=1.
The received signal (\ref{eq:sys_1}) can be further given
\begin{equation} \label{eq:sys_1a}
\begin{split}
y_k=&\bigg(\mathbf{g}_{k}^T \prod_{i=1,\cdots, I_k} \mathbf{\Phi}_i \mathbf{H}_{i}+\mathbf{w}_k \bigg) \mathbf{f}_kx_k + \\
&\sum_{j, j\neq k}^K \bigg(\mathbf{g}_{k}^T \prod_{i=1,\cdots, I_k} \mathbf{\Phi}_i \mathbf{H}_{i}+\mathbf{w}_k \bigg) \mathbf{f}_jx_j+n_k
\end{split}
\end{equation}
where $\mathbf{f}_m$ is the beamforming vector for the $m^{th}, m\neq k$ user.
Furthermore, the SINR at the $k^{th}$ user is written as
\begin{equation} \label{eq:sys_3}
\rho_{k}=\frac{|(\mathbf{g}_{k}^T \prod_{i=1,\cdots, I_k} \mathbf{\Phi}_i \mathbf{H}_{i}+\mathbf{w}_k) \mathbf{f}_k|^2}{|(\sum_{j,j\neq k}^K\mathbf{g}_{k}^T \prod_{i=1,\cdots, I_k} \mathbf{\Phi}_i \mathbf{H}_{i}+\mathbf{w}_k) \mathbf{f}_j|^2+\sigma_n^2}
\end{equation}

\subsection{Problem Formulation}

Our main objective is to combat the propagation attenuations of THz communications by leveraging multi-hop RIS-assisted communication scheme. Therefore, we use the ergodic sum rate as the evaluate metric. However, the major obstacles to maximize the sum rate are to obtain the optimal design of digital beamforming matrix $\mathbf{F}$ and beamforming matrix $\mathbf{\Phi}_{i}, \forall i$, i.e., phase shift matrix of RISs. The optimization problem is formulated as follows,
\begin{equation} \label{eq:BD_1}
\begin{split}
&  \max\limits_{\mathbf{F},\mathbf{\Phi}_i} C(\mathbf{F},\mathbf{\Phi}_{i,\forall i}, \mathbf{w}_{k, \forall k}, \mathbf{g}_{k, \forall k},\mathbf{H}_{i,\forall i})=\sum_{k=1}^K\log_2(1+\rho_k)  \\
& \; \textrm{s.t.} \;\;  tr\{\mathbf{F}\mathbf{F}^{\mathcal{H}} \} \leq P_t \\
& \;\;\;\quad\;\; |\theta_{in_i}|=1\;\forall n_i=1,2,\ldots,N_i.
\end{split}
\end{equation}
Unfortunately, we can easily find that the optimization problem (\ref{eq:BD_1}) is a NP-hard problem because of the non-trivial objective function and the non-convex constraint. As we all know, it is nearly impossible to obtain an analytical solution by the traditional methods of mathematical analysis for the  multi-hop optimization. In addition, exhaustive numerical search is also impractical for large scale networks. Although there are some existing approximation methods that are proposed based on the alternating method to find the sub-optimal solutions for single hop RIS-based system, e.g., \cite{chongwentwc2019,wqq2019beamforming,
Basar2019access}, they are difficult to work for the multi-hop scenario, especially we do not know how many RIS hops the transmitted signal experienced  to arrive $k^{th}$ user, i.e., $I_k (I_k \leq N)$ in prior. Instead, in this paper, we will propose a new method by leveraging the recent advance on DRL technique, rather than directly solving this challenging optimization problem mathematically.

\section {DRL-based Design of Digital and Analog Beamforming}
In this section, we give the details of the proposed DRL-based algorithm for hybrid beamforming of multi-hop THz communication networks utilizing the deep deterministic policy gradient (DDPG) algorithm.

\subsection {Framework of DRL}
Generally, a typical DRL framework consists of six fundamental elements,  i.e., the state set $\mathbf{S}$, the action set $\mathbf{A}$, the instant reward $r(s,a), (s\in \mathbf{S}, a \in \mathbf{A})$, the policy $\pi(s,a)$, transition function $\mathbf{P}$ and Q-function $Q(s,a)$.  Note that the policy $\pi(s,a)$ denotes the conditional probability of taking action $a$ on the instant state $s$. This also means that the policy $\pi(s,a)$  needs to satisfy $\sum_{a \in \mathbf{A}, s \in \mathbf{S},}\pi(s,a) =1$. In addition, since we consider a mobile environment, the transition function $\mathbf{P}$ usually is affected by the environment itself and the action from the RL agent.

\begin{figure}
\begin{center}
  \includegraphics[width=8.5cm]{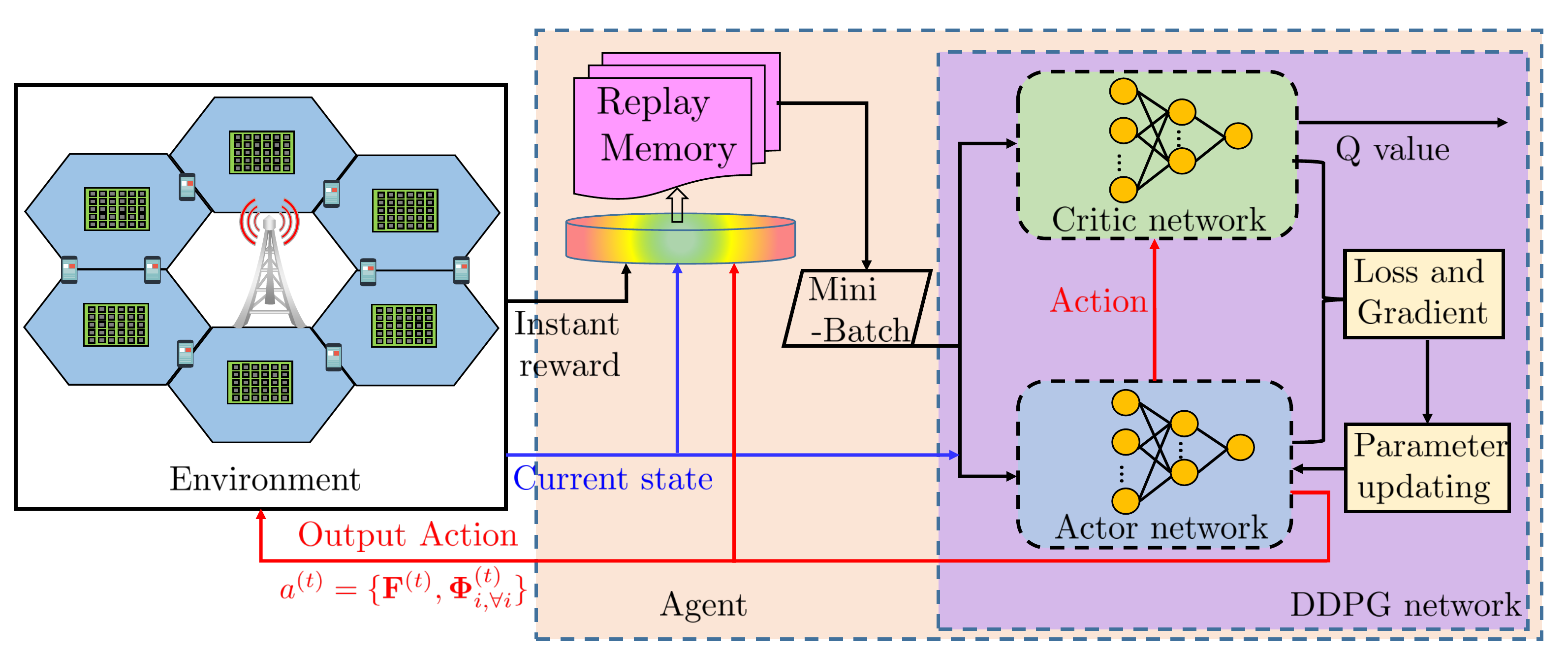}
  \caption{ The illustration of the proposed DRL framework, and the actor-critic DDPG algorithm. }
  \label{fig:NN} \vspace{-8mm}
\end{center}
\end{figure}

Regarding to our proposed hybrid beamforming problem that have an approaching infinite state and action space, the storage size and search complexity of Q-table are extremely impractical. To overcome these issues, we employ a deep Q-learning method to approximate the Q-table by leveraging the universal approximation feature of deep neural networks (DNNs) \cite{universal}. As shown in Fig. \ref{fig:NN}, our proposed DRL framework uses two DNNs (also named actor network and critic network)  to approximate the state/action value function. In other words, a actor neural network to approximate a policy based on the observed environment $s$ state and output an action, while another DNN implements the critic network denoted $Q(\mathbf{\theta} |s(t),a(t))$ to evaluate  the current policy according to the received the rewards.

\subsection{Critic and Actor Networks}
As can be seen in Fig. \ref{fig:NN} that the hardcore of DDPG structure is the critic and actor networks  that are  comprised of a fully connected DNN, where they share the similar structure, consisted of four layers, i.e., two hidden layers with one input and output layer.  Note that the increase or decrease of the width of the network depends on the actions, but the last output layer is up to the number of users. Based on this, we introduce the batch normalization layer between these two hidden layers with ReLU activation function. The optimizer used in the critic and actor networks is Adam with learning rate $\mu_c^{(t)}=\lambda_c\mu_c^{(t-1)}$ and $\mu_a^{(t)}=\lambda_a\mu_a^{(t-1)}$, where $\lambda_c$ and $\lambda_a$ represent their the decaying rate.
\subsubsection{Critic Process}
The main objective of critic agent is to evaluate how good a policy is. The input of the critic network are the current environment state and actions generated by the actor network, and outputs the Q-function based on the DDPG. Its learning rate usually is set as smaller to avoid oscillation, but it needs more time to converge. In order to the negative inputs, the activation function $tanh$  is obtained for the training critic network. To remove the correlation of adjacent states, the input state $\mathbf{S}$ needs the whitening process.

\subsubsection{Actor Process}

The function of the actor network is to learn the current environment based on the DDPG algorithm (Details can be seen in algorithm 1 of our previous work \cite{Ref14c}) and outputs the actions to the critic network. Unlike the critic network, the actor needs additional process, i.e., power modular normalization before the output to implementation problems for computing the  $\Delta_{a}q(\mathbf{\theta}_c^{(target)}|s^{(t)},a)$.  The approximate policy gradient might yield the error, and it also could not make sure that we can obtain the optimal solution, but we can minimize the error by using the compatible  features of transition function.

In the actor-critic RL agent, the policy parameters and transition function are updated simultaneously, and $\mathbf{F}$ needs to meet the  power constraint $Tr \{\mathbf{F}\mathbf{F}^{\mathcal{H}} \} =P_t$. In order to satisfy this condition, a normalization layer is added at the output of the actor network.  Noting that, the signal is changed the transmission direction after it goes through RISs, but its amplitude be maintained as $|\mathbf{\Phi}_i(n_i,n_i)|^2=1$ since it does not consume the additional power.

\subsection{Proposed DRL Algorithm }

Before we implement the proposed DRL algorithm, the channel information, $\mathbf{H}_i, i=1,\cdots, I$, $\mathbf{w}_{k}$ and $\mathbf{g}_{k} \forall k$ are collected by the existing methods that are investigated by some previous works \cite{ning2019beamforming,weili_CE_2020}. The channel information and previous actions of $\mathbf{F}^{(t-1)}$ and $\mathbf{\Phi}_i^{(t-1)}, \forall i$ at previous $t-1$ state, the agent obtains the current state $s^{(t)}$. In addition, weight initialization is also a key factor to affect the learning process. The action $\mathbf{F}$ and $\mathbf{\Phi}_i, \forall i$, networks parameters $\mathbf{\theta}_c^{(train)}$, $\mathbf{\theta}_a^{(train)}$,  $\mathbf{\theta}_c^{(target)}$,  and $\mathbf{\theta}_a^{(target)}$and replay buffer $\mathcal{M}$ should be initialized before running the algorithm. Furthermore, we also proposed two initialization algorithms, one is based singular value decomposition (SVD), while another one is utilize the max-min SINR method.

The algorithm stops when it converges or reaches the maximum number of iteration steps. The obtained rewards could not be increased with the taking more actions, we think that the output $\mathbf{F}_{opt}, \mathbf{\Phi}_{i,opt}$  are the optimal. Noting that the proposed algorithm might converges the sun-optimal solutions although our objective is to obtain the optimal digital  and analog beamformings. Combing with the previous stated DDPG, the whole proposed DRL-algorithm can be summarized as Algorithm \ref{alg:ALG1} in the following.

\begin{algorithm}[H]
\caption{DRL-based hybrid beamforming design for RIS-based THz Systems}
\label{alg:ALG1}
\textbf{Input:} $\mathbf{w}_{k,\forall k}$, $\mathbf{g}_{k, \forall k},\mathbf{H}_{i,\forall i}$ \\
\textbf{Output:} The optimal $a=\{ \mathbf{F},\mathbf{\Phi}_{i,\forall i} \}$, $Q$ function\\
\textbf{Initialization:} Memory $\mathcal{M}$; parameters $\mathbf{\theta}_c^{(train)}$, $\mathbf{\theta}_a^{(train)}$, $\mathbf{\theta}_c^{(target)}$, $\mathbf{\theta}_a^{(target)}$; \\ beamforming matrices $\mathbf{F}$, $\mathbf{\Phi}_{i,\forall i} $
\begin{algorithmic}[1]
\WHILE{}
\FOR{\texttt{espisode $=0,1,2, \cdots,Z-1$}}
\STATE Collect and preprocess $\mathbf{w}_{k, \forall k}^{(n)}, \mathbf{g}_{k, \forall k}^{(n)},\mathbf{H}_{i,\forall i}^{(n)}$ for the $n^{th}$ episode to obtain the first state $s^{(0)}$
\FOR{\texttt{t=$0,1,2, \cdots, T-1$}}
\STATE Update action $a^{(t)}=\{\mathbf{F}^{(t)},\mathbf{\Phi}_{i,\forall i}^{(t)} \}=\pi(\mathbf{\theta}_a^{(train)})$ from the actor network  \\
\STATE Implement  DDPG Algorithm \\
\STATE Update parameters $\mathbf{\theta}_c^{(train)}$, $\mathbf{\theta}_a^{(train)}$, $\mathbf{\theta}_c^{(target)}$, $\mathbf{\theta}_a^{(target)}$ \\
\STATE Input them to the agent as next state  $s^{(t+1)}$
\ENDFOR
\ENDFOR \\
\textbf{Until:} Convergent or reaches the maximum  iterations.
\ENDWHILE
\end{algorithmic}
\end{algorithm} \vspace{-4mm}

In terms of this proposed algorithm, its state, action, reward and convergence are elaborated in the following.

\subsubsection{State}
The state $s^{(t)}$ is continuous and constructed by the transmit digital beamforming matrix $\mathbf{F}^{(t-1)}$, the analog beamforming matrices $\mathbf{\Phi}_i^{(t-1)}, \forall i$ in the previous $t-1$ time step, and the channel information $\mathbf{H}_i, i=1, \cdots, I$, $\mathbf{w}_k$ and $\mathbf{g}_k, \forall k$. Since the DRL based on the TensorFlow platform do not support the complex number inputs, we employ two independent input ports to input the real part and the imaginary part of state $s$ separately. We have the dimension of the state space $D_s=2MK+2\sum_{i=1,\cdots, I}N_i+2MN_1+2\sum_{i=1, \cdots, I-1}N_iN_{i+1}+2KN_I$. We assume that there is no neighboring interference between the different states. To maximize the transmission distance, assume that each state can offer some prior knowledge to DRL agent for selecting the optimal RIS and analog beamforming. The optimal beamforming is related to the channel information and interference to other users. Then, DRL agent can learn the interference patten from the historical date, so that it can infer the future interference at the next time step.

\subsubsection{Action}
Similarly, the action space is also continuous, and  comprised of the digital beamforming matrix $\mathbf{F}$ and analog beamforming matrices  $\mathbf{\Phi}_i, \forall i$. Furthermore, the real and imaginary part of $\mathbf{F}=Re\{\mathbf{F} \}+Im \{ \mathbf{F}\}$ and $\mathbf{\Phi}_i=Re \{\mathbf{\Phi}_i \} +Im \{\mathbf{\Phi}_i \}$ are also separated as two inputs. Its dimension also depends the parameters of communication systems, as $D_a=2MK+2\sum_{i=1,\cdots, I}N_i$.


\subsubsection{Reward}
The instant rewards is affected by two main factors: the contributions to throughput $C(\mathbf{F}^{(t)},\mathbf{\Phi}_i^{(t)}, \mathbf{w}_{k}, \mathbf{g}_{k},\mathbf{H}_{i=1,\cdots, I})$ and the penalty caused by the adjusting the beamforming direction under the prior information, the instantaneous channels $\mathbf{H}_{i=1,\cdots, I}$, $\mathbf{h}_{k}, \forall k$ and the actions  $\mathbf{F}^{(t)}$ and $\mathbf{\Phi}_i^{(t)}$ outputted  from the actor network.

\subsubsection{Convergence}

Furthermore, there are some factors that can affect the convergence. For example, the initialization of action and state parameters plays a key role, which will introduced in the following. In addition, gradient evolution, learning rate also pose the affect on convergence. The too large or small gradient and learning rate both make a algorithm diverge. We investigate the affect of the learning rate that are shown in simulation section.

\section{Numerical Results}

In this section, we numerically evaluate the performance of the proposed DRL based algorithm for DRL-based hybrid beamforming for multi-hop multiuser RIS-assisted wireless THz communication networks.

\subsection{Simulation Settings}
In the following simulations, we consider a single cell scenario, where there is only one BS, and many RISs that are randomly deployed in a circular region with the diameter as 100 m.
\subsubsection{System Model}
We employ the proposed hybrid beamforming architecture shown in  Fig. \ref{fig:hybrid}. In particular, The BS has $M=8$ antennas with the same number of RF chains, and $K=32$ mobile users equipped the single antenna and RF chain. To reduce the complexity of deployment and learning, we adopt that all $N=64$ RISs have the same number of elements, i.e., $N_i=128$ for all $i$, and the spacing between
elements equal to $2\lambda$. The channel matrices $\mathbf{w}_{k, \forall k},\mathbf{g}_{k, \forall k},\mathbf{H}_{i-1,\forall i}$ are generated randomly with Rayleigh distribution in the simulations. The transmission frequency is set as 0.12 THz occupied the fixed 12 GHz bandwidth, and the transmitted power of BS is set as $10$ Watt.

\subsubsection{DRL Settings}

Without special highlight, the parameter settings of the proposed DRL-based beamforming algorithm are concluded in Table \ref{tab:hyperP}.
\begin{table}
\caption{ Parameters for DRL-based beamforming algorithm} \label{tab:hyperP}
\begin{center} \vspace{-4mm}
\begin{tabular}{ | m{4em} | m{19em}| m{4em} | }
\hline
Parameters & Description &  Settings \vspace{1mm}\\
\hline
$\beta$ & Discounted rate of the future reward & 0.99  \vspace{1mm}\\
\hline
$\mu_c$ & Learning rate of training critic network update & 0.001  \vspace{1mm} \\
\hline
$\mu_a$ & Learning rate of training actor network update & 0.001 \vspace{1mm} \\
\hline
$\tau_c$ & Learning rate of target critic network update & 0.001 \vspace{1mm} \\
\hline
$\tau_a$ & Learning rate of target actor network update & 0.001 \vspace{1mm} \\
\hline
$\lambda_c$ & Decaying rate of training critic network update & 0.005  \vspace{1mm}\\
\hline
$\lambda_a$ & Decaying rate of training actor network update & 0.005 \vspace{1mm} \\
\hline
$D$ & Buffer size for experience replay& 100000 \vspace{1mm} \\
\hline
$Z$ & Number of episodes & 5000 \vspace{1mm} \\
\hline
$T$ & Number of steps in each episode & 20000 \vspace{1mm} \\
\hline
$W$ & Number of experiences in the mini-batch & 16 \vspace{1mm} \\
\hline
$U$ & Number of steps synchronizing target network with the training network & 1 \vspace{1mm} \\
\hline
\end{tabular} \vspace{-4mm}
\end{center}
\end{table}

\subsubsection{Benchmarks}

To show the effectiveness of our proposed, three significant cases are selected as benchmarks. The first case is an ideal case, where there is no RISs to assist transmit, i.e., $I=0$, and we employ the full digital zero-forcing beamforming. The second typical benchmark was already investigated in some existing works \cite{chongwentwc2019,ning2019beamforming,Nie2019resource,wqq2019beamforming}, where there is a just single hop between the BS and each user, and  an alternating optimization method is usually proposed to design the beamforming matrices.

\subsection{Comparisons with Benchmarks}
\begin{figure}[tb]  \vspace{-4mm}
\begin{center}
  \includegraphics[width=8.2cm]{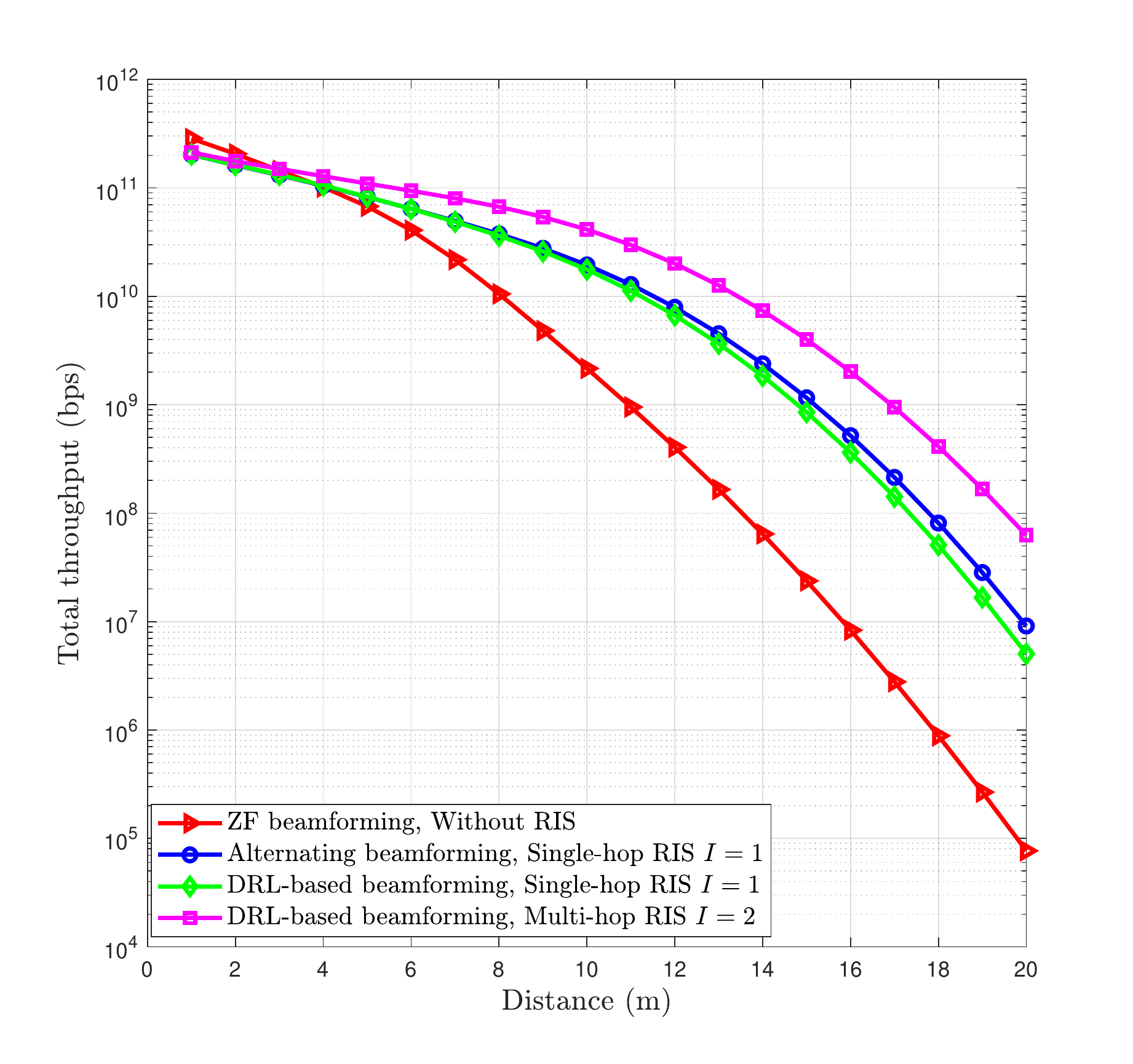} \vspace{-4mm}
  \caption{Total throughput versus transmission distance. We compare  the performance of four schemes. } 
  \label{fig:comparison} 
\end{center} \vspace{-8mm}
\end{figure}

We compare the proposed DRL-based method described in Algorithm 1 for multi-hop RIS-assisted wireless THz communication networks as well as three mentioned benchmarks shown in  Fig. \ref{fig:comparison}. It shows that the proposed DRL-based multi-hop (i.e., $I=2$) THz communication scheme nearly always obtain the best system throughput compared with the considered three schemes over the whole transmission distance from 1m to 20m. In particular, we employ the ideally full digital ZF beamforming for the first benchmark, where does not have the RIS to assist transmission, its throughput drops  fastest with the increase of the transmission distance. For example, under the same throughput 1Gbps, we can see that the proposed DRL-based two-hop scheme obtains around 50\% and 14\% more transmission distances than that of ZF beamforming without RIS and single-hop scheme respectively. What's more, this performance gap will becomes larger when the transmission distance increases. Another interesting point is that the traditional alternating-based method that we adopt is the proposed method in \cite{wqq2019beamforming}, as this benchmark can obtain a little better performance than that of the DRL-based beamforming single-hop scheme, but much less than that of the two-hop scheme.

\subsection{Impact of System Settings}

\begin{figure}[htbp] \vspace{-4mm}
\begin{center}\vspace{-0mm}
  \includegraphics[width=8.2cm]{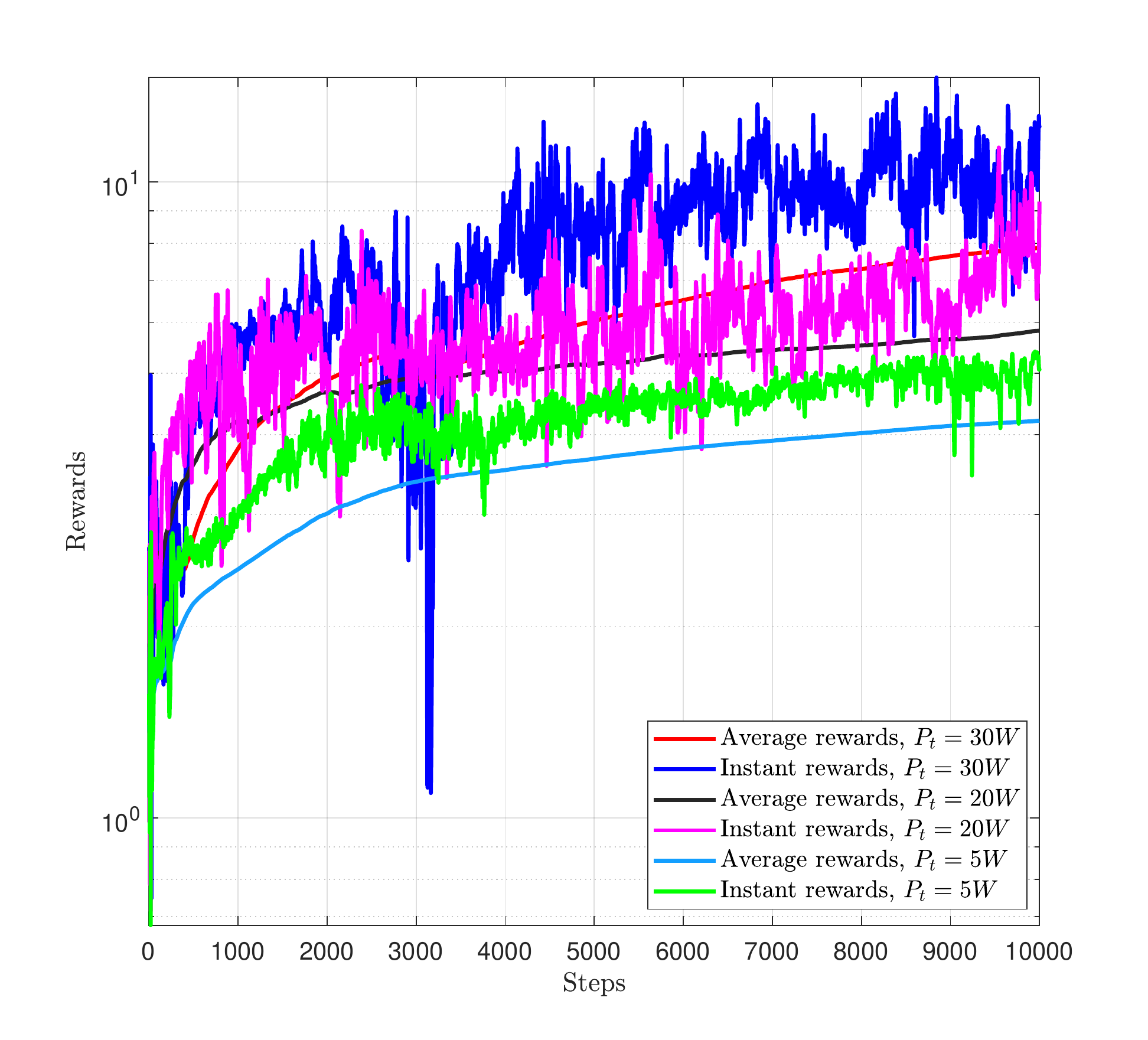} \vspace{-4mm}
  \caption{Rewards versus steps at $P_t=5W$, $P_t=20W$, and $P_t=30W$ respectively. }
  \label{fig:steps5dB} \vspace{-8mm}
\end{center}
\end{figure}

To further verify the effectiveness of our proposed scheme, we have evaluated its rewards performance as a function of time steps,  which is shown in Fig. \ref{fig:steps5dB}, where we consider the setting of $M=8,I=2, N_i=64, K=4$.  It can be seen that, the sum rate converges with time step $t$. With the increasing of SNR, instant and average rewards both increase naturally. However, it converges faster at the low transmission power $P_t=5W$ than that of high transmission power $P_t=30W$. This is because the higher transmission power will means the state spaces of instant rewards are larger, which needs to more time to converge the local optimal solution. Based on these results, we also conclude that our proposed DRL-based algorithm can learn from the environment and feed the rewards to the agent to prompt the beamforming matrices $\mathbf{F}$ and $\mathbf{\Phi}_{i,\forall i} $ converging the local optimal.

\section{Conclusions}
In this paper, a novel and practical hybrid beamforming architecture for multi-hop multiuser RIS-assisted wireless THz communication networks was proposed, which can effectively combat the severe propagation attenuations and improve the coverage range. Based on this proposed scheme,  a non-convex joint design problem of the digital beamforming and analog beamforming matrices was formulated.  To tackle this NP-hard problem,  a novel DRL-based algorithm was proposed, which is a very
early attempt to address this hybrid design problem. Simulation results show that our proposed scheme is able to improve 50\% more coverage range of THz communications compared with the considered benchmarks. Furthermore, it is also shown that our proposed DRL-based method is a state-of-the-art method to solve the NP-bard beamforming problem, especially when the signals at RIS-assisted THz communication networks experience multi hops.

\vspace{-4mm}
\bibliographystyle{ieeetran}   
\bibliography{multihop_ris} 

\end{document}